\documentclass[graybox]{svmult}

\usepackage{type1cm}        
\usepackage{makeidx}         
\usepackage{graphicx}        
\usepackage{multicol}        
\usepackage[bottom]{footmisc}

\usepackage{newtxtext}       %
\usepackage{newtxmath}       
\usepackage{lineno}
\usepackage{amsmath}
\usepackage{url}
\usepackage{subeqnarray,amsmath,amsfonts,graphicx}

\newcommand{\fo}{\rm}

\def\dfrac{\displaystyle\frac}

\def\dint{\displaystyle\int}

\def\2q{\quad\quad}
\def\3q{\quad\quad\quad}

\def\<{\left\langle}
\def\>{\right\rangle}

\def\:{{\,:\,}}

\def\tot1{\sigma_{ij}^{(1,T)}}
\def\tot2{\sigma_{ij}^{2,T}}
\def\tot{\sigma_{ij}}

\makeindex             

\begin{document}

\title*{An SEIR epidemic model of fractional order  
to analyze the evolution of the  COVID-19 epidemic in Argentina}
\author{Juan E. Santos       
\and Jos\'e M. Carcione 
\and Gabriela B. Savioli 
\and Patricia M. Gauzellino}

\institute{ Juan E. Santos
\at
School of Earth Sciences and Engineering, Hohai University, Nanjing 211100, China.  
\\ Universidad de Buenos Aires, Facultad de Ingenier\'\i a, Instituto del Gas y del Petr\'oleo, Av. Las Heras 2214, Piso 3, Buenos Aires, Argentina.  \\
Departmet of Mathematics, Purdue University. United States. \\
              Tel.: +54-11-52850341. 
              \email{santos@purdue.edu}           
\and
Jos\'e M. Carcione 
\at
National Institute of Oceanography and Applied Geophysics - OGS, Trieste, Italy. \\
School of Earth Sciences and Engineering, Hohai University, Nanjing 211100, China.  
\and
Gabriela B. Savioli
\at Universidad de Buenos Aires, Facultad de Ingenier\'\i a, Instituto del Gas y del Petr\'oleo, Av. Las Heras 2214, Piso 3, Buenos Aires, Argentina
\and
Patricia M. Gauzellino
\at
Facultad de Ciencias Astron\'omicas y Geof\'\i sicas, Universidad
Nacional de La Plata, Argentina
}

%
%
\maketitle

\abstract*{Each chapter should be preceded by an abstract (no more than 200 words) that summarizes the content. The abstract will appear \textit{online} at \url{www.SpringerLink.com} and be available with unrestricted access. This allows unregistered users to read the abstract as a teaser for the complete chapter.
Please use the 'starred' version of the \texttt{abstract} command for typesetting the text of the online abstracts (cf. source file of this chapter template \texttt{abstract}) and include them with the source files of your manuscript. Use the plain \texttt{abstract} command if the abstract is also to appear in the printed version of the book.}

\abstract{A pandemic caused by a new coronavirus (COVID-19) has spread worldwide, inducing  an epidemic still active in Argentina.  
In this chapter, we present a case study using  an  SEIR (Susceptible-Exposed-Infected-Recovered) diffusion model 
of fractional order in time  to  analyze  
the evolution of the epidemic  in  Buenos Aires and neighboring areas (Regi\'on Metropolitana de Buenos Aires, (RMBA)) comprising about   15 million inhabitants. In the  SEIR model,  individuals are divided into 
four classes, namely, susceptible (S), exposed (E),  infected (I) and 
recovered (R).
The SEIR model of fractional order allows for the  incorporation of memory, with hereditary properties of the system, being a generalization of the classic SEIR first-order system,  where such effects are ignored. Furthermore,  the fractional 
model  provides one additional parameter  to obtain a better fit of the  data. 
The parameters of the model are calibrated by using as data the number of casualties officially reported. 
Since infinite solutions honour the data,  
we show a set of cases with different values of  the lockdown parameters, 
fatality rate, and incubation and infectious periods. 
The different reproduction ratios $R_0$ and infection fatality rates 
(IFR) so obtained  indicate the results may differ from  recent
reported values, constituting possible alternative solutions. A comparison with 
results obtained with the classic SEIR model is also included. The  analysis allows us to study  how isolation and social distancing measures affect the time evolution of the epidemic.}

\section{Introduction}\label{intro}

We present an SEIR subdiffusion  model  
of fractional order $\nu$, with $0 <\nu \le 1$  to analyze the time evolution of the COVID-19  epidemic in  Buenos Aires and neighboring areas (Region Metropolitana de Buenos Aires, (RMBA)) with a population of about 15 million inhabitants. 
{\fo RMBA consists  of Ciudad Aut\'onoma de Buenos Aires (CABA)  plus forty municipalities 
 covering an area of about thirteen thousand square kilometers, where some of these municipalities have rural  areas. Thus, RMBA has an average  population density 
 of  1100 people/km$^2$, but  in CABA and many of its neighboring cities this number increases significantly. For example, CABA has a population density of  about 14000 people/km$^2$.
In this work, we  consider that RMBA has a  uniform population distribution.}

The epidemic  started officially on March 9th with the number of cases and deaths  
still increasing {\fo at the day of writing (September 22th, 2020}.
The classical  SEIR model ($\nu = 1$)  has been used by Carcione et al. \cite{carcione2020} and Santos et al. \cite{Santos2020} to model the COVID-19  epidemic in Italy and Argentina, respectively.   

Fractional calculus has been used to  define diffusion and wave propagation  models  in biological  and  viscoelastic materials \cite{caputo67,mainardi96,carcione2002,mainardi2010,caputo2011,caputo2011b,kochubei2011,carcione2017}.   
One important
property  of the fractional-order SEIR model  is that   incorporates    memory and 
 hereditary properties, a behavior  exhibited by  most biological systems. 
 {\fo The use of fractional order derivatives affects the 
duration of the epidemic, peaks of infected and dead individuals per day and number of number casualties.}

 Among other authors that have applied   fractional calculus to obtain solutions of the SEIR model,  
 we mention 
 Scherer et al. \cite{scherer2011}, that used a  Gr\"unwald-Letnikov  time-discrete  procedure, introduced by   
 Ciesielski and Leszczynski \cite{cl2003} (CL method). 
 Besides, 
 Zeb et al. \cite{zeb2013} presented  an analysis of several numerical methods to solve the SEIR model 
 of fractional order.  For  general works on fractional calculus including numerical methods,  
 we refer to Podlubny \cite{podlubny99} and Li and Zeng \cite{lizeng2015}.
 
We first  formulate an initial-value problem (IVP) for the  {\fo classical SEIR model ($\nu = 1$)} and the SEIR subdiffusion equations  of 
 fractional order $\nu$   at the continuous level  using the 
 Caputo definition of the fractional derivative \cite{mainardi2010}. 
 Existence and uniqueness of the solution of this IVP, with positive values,  is demonstrated  in    \cite{zeb2013}.
 The numerical solutions of the  
 continuous IVP  are computed  by using the  time-explicit  
 algorithm of  Gorenflo-Mainardi-Moretti-Paradisi (GMMP method)
 \cite{gorenflo2002,gorenflo2007}.  
 The conditional stability of the time-explicit GMMP method (and also of the CL method) was demonstrated by 
 Murillo et al. \cite{murillo2017} [see their equation (19)].
 The validation of the   GMMP method  is performed by comparison of its  results against those of the classic  SEIR model  and those of the  fractional Adams-Bashford-Moulton  method (ABM method) as defined in \cite{lizeng2015}. 
 
The parameters of the SEIR model  are  the  birth and  death  rates, 
infection and incubation periods, probability of disease transmission
per contact, fatality rate and initial number of exposed 
individuals. These parameters, together with the 
order of the fractional  derivative,  are obtained by fitting the number of fatalities  officially reported. 
This  is   an inverse problem with an infinite number of solutions (local minima) honouring the data,  which  is solved by using  a quasi-Newton  technique for nonlinear least squares problem with the formula  of Broyden-Fletcher-Goldfarb-Shanno \cite{Gill81}.
The numerical simulations give an effective procedure  to study the spread of the evolution of virus, analyze the effects of the 
lockdown  measures and predict the peak of infected and dead individuals {\fo per day}.

\section{The Caputo derivative and initial value problems}\label{sec:1}

 For $0 < \nu \le 1$, the time fractional Caputo  derivative $D^{\nu}_c (u(t)$  is defined as \cite{caputo67,gorenflo2002,gorenflo2007,mainardi2010}
 \begin{eqnarray}\label{defcaputo}
 D^{\nu}_c (f(t) 
 =  \dfrac{1}{\Gamma(1 - \nu)} 
 \dint_0^t\left[ \dfrac{\partial}{\partial f(\tau)} \right]
 \dfrac{ d \tau}{(t - \tau)^\nu}, 
 \end{eqnarray}
 where $\Gamma(\cdot )$ denotes the  Euler's  Gamma function.
 
 Note that the Caputo derivatives of  constant functions $f(t) = 1$ vanish and those of  powers of $t$, $f(t) = t^k$  are 
\[
\dfrac{\Gamma(k+1)}{\Gamma(k - \nu +1)} t^{k - \nu}.
\]
The advantage of using the Caputo derivative in Caputo-type IVP's  is that the initial  conditions are the same as those of the classical
 ordinary differential equations. For details on the Caputo derivative and its relation with the 
 Riemann-Liouville fractional derivative we refer to \cite{mainardi2010}.
 
To  approximate  the time-fractional Caputo derivative, we use a backward  Gr\"unwald-Letnikov  
approximation  at time $t_n = n \Delta t, n= 0,1,,\cdots$, with $f_n = f(n \Delta t)$, $\Delta t $ being 
the time step,   as follows 
\cite{gorenflo2002,gorenflo2007}:

\begin{eqnarray}\label{eq2}
D^{\nu}_c (f(t)|_{t_{n+1}} \approx \dfrac{1}{(\Delta t)^{\nu}}\sum_{j=0}^{n+1} (-1)^j 
c^\nu_j \binom  \nu  j f_{n+1-j}.
\end{eqnarray}

The coefficients
\[
c^\nu_j = (-1)^j  \binom  \nu  j 
\]
can be  obtained in terms of Euler's Gamma function  using the recurrence relation 
\begin{eqnarray}\label{eq2a}
&&\binom  \nu  j =  \dfrac{\Gamma(\nu+1)}{\Gamma(j+1) \Gamma(\nu-j+1)} = \dfrac{\nu-j+1}{j} 
\binom  \nu {j-1}, \quad  \binom  \nu  0 = 1.
\end{eqnarray}
The work by Abdullah et al. \cite{abdullah2017} presents an 
analysis of the fractional-order SEIR  model formulated in terms of the Caputo derivative and 
its  GMMP time discretization. 

\section{The classical and fractional-order SEIR  models}\label{seir}

The  IVP for the classic SEIR  system  of  nonlinear ordinary differential equations is  
\begin{eqnarray}\label{eq4}
&&\dot S =  f_1(S,E,I,R)(t) = \Lambda - \mu S(t) - \beta S(t) \dfrac{I(t)}{N(t)},\\ 
&&\dot E =  f_2(S,E,I,R)(t) = \beta S(t) \dfrac{I(t)}{N(t)} -  (\mu + \epsilon) E(t), \nonumber\\ 
&&\dot I =  f_3(S,E,I,R)(t) = \epsilon E(t)  - (\gamma + \mu + \alpha) I(t), \nonumber \\ 
&&\dot R =  f_4(S,E,I,R)(t) = \gamma I(t) - \mu R(t),  \nonumber
\end{eqnarray}
with initial conditions $S(0), E(0), I(0)$ and $R(0)$.  A  dot  above a variable indicates the time derivative, while   $N(t)$ is the number of live 
individuals at time $t$, i.e., $N = S + E + I + R \le N_0$, {\fo $N_0$ being the total initial population.} 
In  \eqref{eq4},   
 $S$ is the  number of individuals 
susceptible to be exposed  while  $E$ is the number of  exposed individuals, in which the disease is latent; they are infected but not infectious.  Individuals in the  $E$-class 
become infected ($I$) with a rate $\epsilon$ and infected become recovered ($R$) with a rate $\gamma$.  
People in the $R$ class do not move back to the $S$ class since lifelong immunity is assumed.
Furthermore,  $1/\gamma$  and $1/\epsilon$ are   
the infection and incubation periods, respectively, $\Lambda$ is the birth rate, $\mu$ is 
the  natural per capita death rate, $\alpha$ is the  average fatality rate, and  $\beta$ is the probability of disease transmission per contact.  
All of these coefficients have units  of 1/time. 
{\fo 
Given the short period of the epidemic in Argentina (6 months at the time of writing), 
and that the average life expectancy
is about 76 years, it is reasonable to assume that 
$ \Lambda = \mu N$, so that the deaths balance  the newborns.} 

Dead individuals $D(t)$ are computed as $D(t) = N_0 - N(t)$, so that   
the dead people per unit time $\dot D (t)$, can be obtained as \cite{Sen17}:
\begin{equation} \label{eq3}
\dot D (t) = \alpha I (t). 
\end{equation} 

Next, we reformulate the system \eqref{eq4} into a fractional-order system by using the Caputo derivative in \eqref{defcaputo}:

\begin{eqnarray}
&&D^{\nu}_c S(t) = f^\nu_1(S,E,I,R)(t) =\mu^\nu N  - \mu^\nu  S(t) - \beta^\nu  S(t) \dfrac{I(t)}{N(t)},\nonumber\\ 
&&D^{\nu}_c E(t) = f^\nu_2(S,E,I,R)(t) = \beta^\nu S(t) \dfrac{I(t)}{N(t)} 
-  (\mu^\nu + \epsilon^\nu) E(t)\label{eq5}\\
&&D^{\nu}_c I(t) = f^\nu_3(S,E,I,R)(t) = \epsilon^\nu E(t)  - (\gamma^\nu + \mu^\nu + \alpha^\nu) I(t), \nonumber\ \\ 
&&D^{\nu}_c R(t)= f^\nu_4(S,E,I,R)(t) =  \gamma^\nu I(t) - \mu^\nu R(t).  \nonumber\
\end{eqnarray}

The  reproduction ratio, $R_0$, indicates   
 the number of  cases induced by a single infectious individual. When $R_0 < 1$,  the disease dies out; when $R_0 > 1$, an epidemic occurs. Al-Sheikh  \cite{alsheikh2012}   analyzes the behavior of the SEIR models 
in terms of  $R_0$. For the SEIR model, $R_0$ is given by \cite{zhang2013}  
\begin{equation} \label{31}
R_0 = \frac{\beta^\nu \epsilon^\nu}{(\epsilon^\nu+\mu^\nu) (\gamma^\nu+ \alpha^\nu +
\mu^\nu)}. 
\end{equation}

The infection fatality rate (IFR) is defined as  
%

\begin{equation} \label{IFR1}
{\rm IFR} \ (\%) = 100 \cdot \frac{\alpha^\nu}{\alpha^\nu + \gamma^\nu} \approx  
100 \cdot \frac{\alpha^\nu}{\gamma^\nu} , 
\end{equation}
where this relation holds at all times, not only at the end of the epidemic. 

\subsection{Time discretization}

An explicit conditionally stable GMMP algorithm for the fractional order system \eqref{eq5} is formulated as follows 
\cite{gorenflo2002,gorenflo2007}:
\begin{eqnarray}
&&S_{n+1} = -\sum_{j=1}^{m+1} c^\nu_j S(m+1 - j) + S_0 \sum_{j=0}^{m+1} c^\nu_j 
+ (\Delta t)^\nu f_1(S_n, E_n, I_n, R_n)\label{gmmp1}\\
&&E_{n+1} = -\sum_{j=1}^{m+1} c^\nu_j E(m+1 - j) + E_0 \sum_{j=0}^{m+1} c^\nu_j 
+ (\Delta t)^\nu f_2(S_n, E_n, I_n, R_n)\label{gmmp2}\\
&&I_{n+1} = -\sum_{j=1}^{m+1} c^\nu_j I(m+1 - j) + I_0 \sum_{j=0}^{m+1} c^\nu_j 
+ (\Delta t)^\nu f_3(S_n, E_n, I_n, R_n)\label{gmmp3}\\
&&R_{n+1} = -\sum_{j=1}^{m+1} c^\nu_j R(m+1 - j) + R_0 \sum_{j=0}^{m+1} c^\nu_j 
+ (\Delta t)^\nu f_4(S_n, E_n, I_n, R_n)\label{gmmp4}
\end{eqnarray}
The results of the GMMP method \eqref{gmmp1}-\eqref{gmmp4} will be validated against the solution of the classical SEIR model ($\nu = 1$) and the Adams-Bashford-Moulton (ABM) time-explicit scheme as defined in 
\cite{lizeng2015} and included in the Appendix.

\section{Numerical results.}

\subsection{Validation of the GMMP algorithm} \label{vali}

 The results of the GMMP algorithm are cross-checked with those of the ABM solver for the 
 classical SEIR model ($\nu = 1$ ) and SEIR  models of fractional orders $\nu = 0.9$ and  $0.8$.
 
 We use the following  parameters,  given in Chowel et al. \cite{chowel2003} and used by Carcione et al. \cite{carcione2020} to perform a parametric analysis of the model.  Average disease incubation $1/\epsilon =3 $ days, 
 infectious period $1/\gamma= 8$ days, induced fatality rate $\alpha= 0.006$/day, $\beta = 0.75$/day, and $\Lambda = \mu = 0$.  The initial conditions are 
 $E(0) = 1, S(0) = N(0) - E(0) - I(0), I(0) = 1$ and $R(0) = 0$. 
 The time step is  $dt$ = 0.01 day and 
 N$_0$ = 10 million. This case corresponds to a high reproduction ratio $R_0 = 5.72$.
 
\begin{figure}
\includegraphics[scale=0.35]{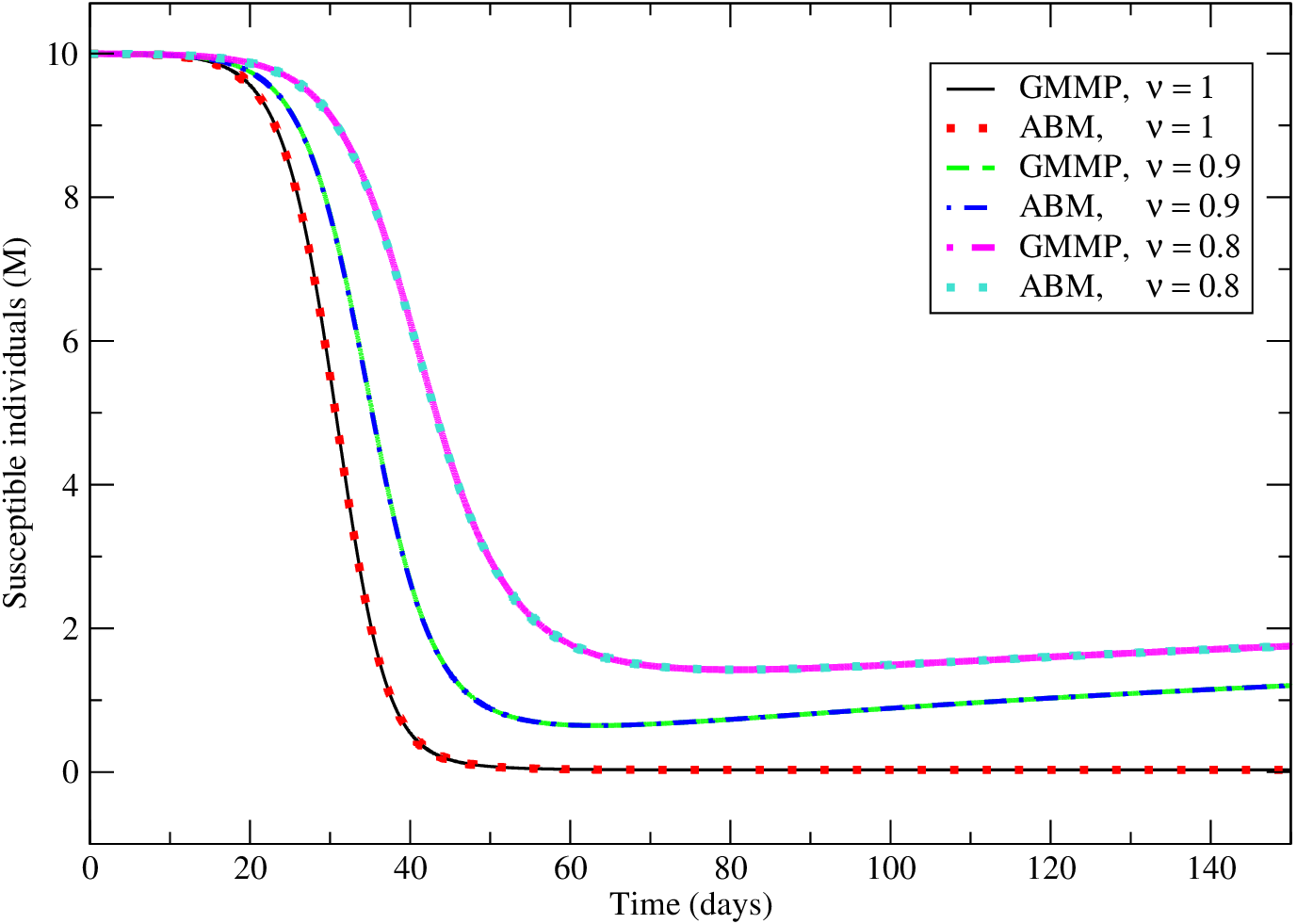}
\vskip0.3cm
\caption{Susceptible individuals for the classical SEIR model ($\nu = 1$) and fractional-order
derivatives $\nu = 0.8$ and  $0.9$}
\label{fig1}       
\end{figure}

\begin{figure}
\vskip1cm
\includegraphics[scale=0.35]{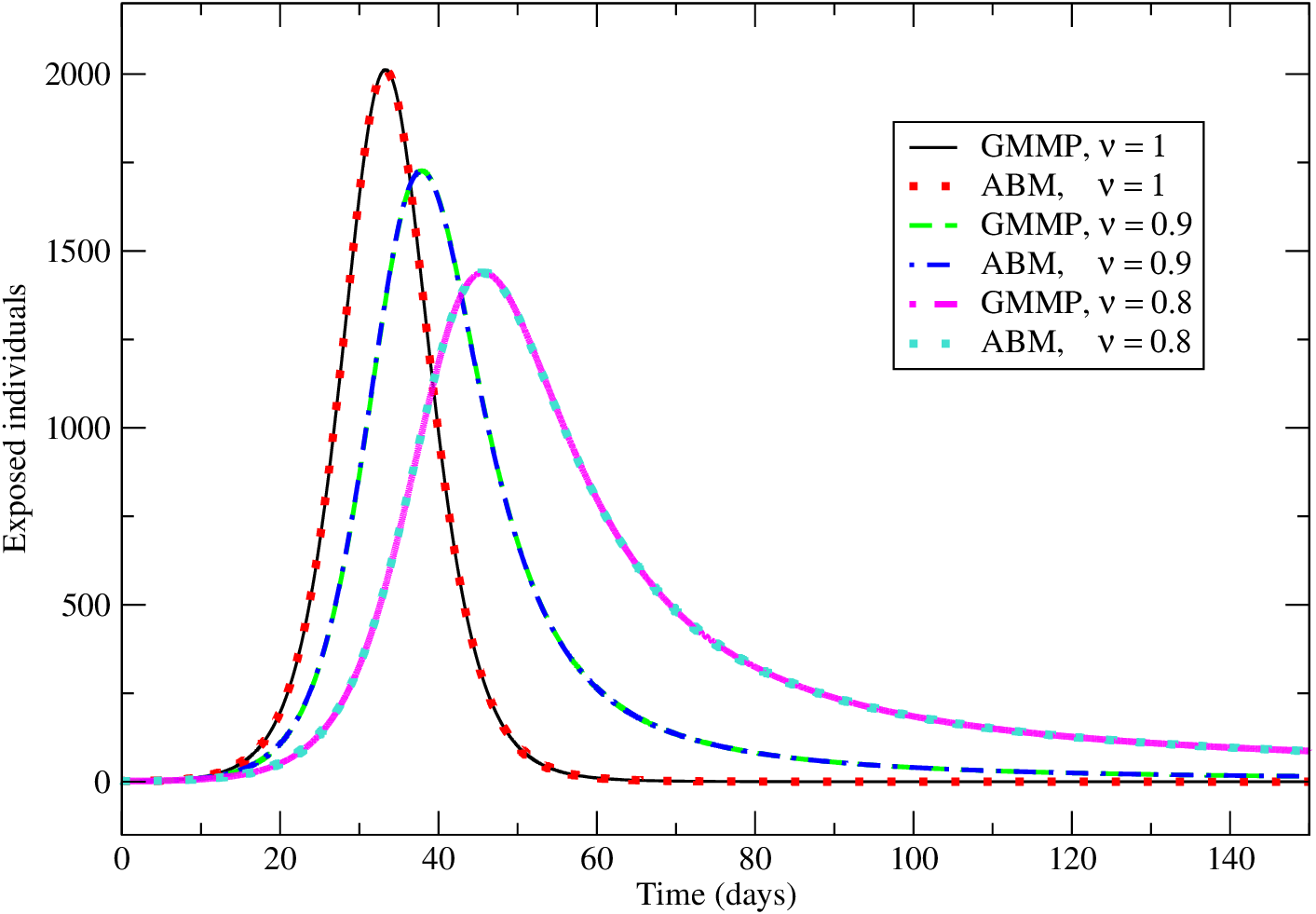}
\vskip0.3cm
\caption{Exposed individuals for the classical SEIR model ($\nu = 1$) and fractional-order
derivatives  $\nu = 0.8$ and  $0.9$}
\label{fig2}       
\end{figure}

\begin{figure}
\vskip1cm
\includegraphics[scale=0.35]{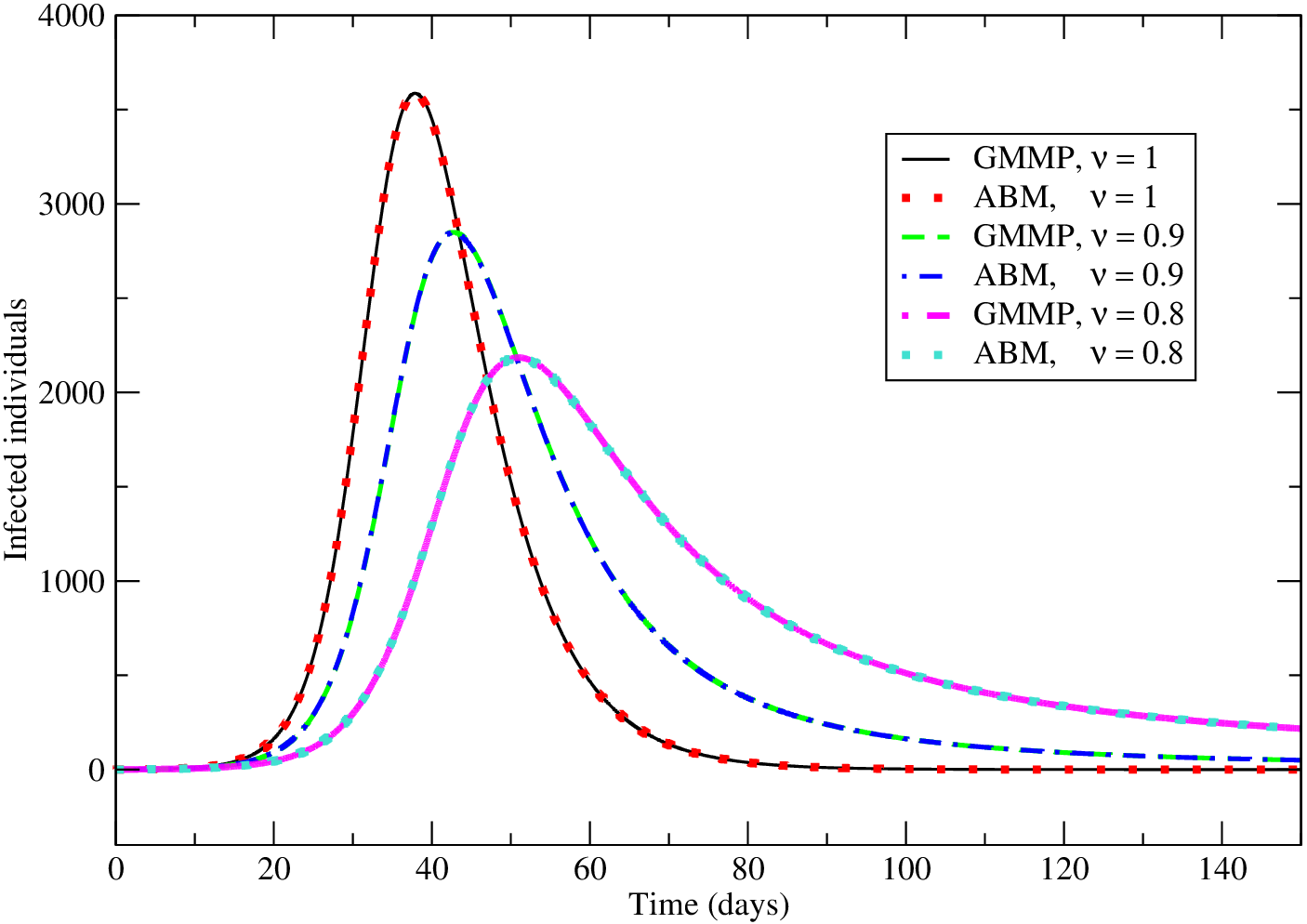}
\vskip0.3cm
\caption{Infected individuals for the classical SEIR model ($\nu = 1$) and fractional-order
derivatives  $\nu = 0.8$ and  $0.9$}
\label{fig3}       
\end{figure}

\begin{figure}
\vskip1cm
\includegraphics[scale=0.35]{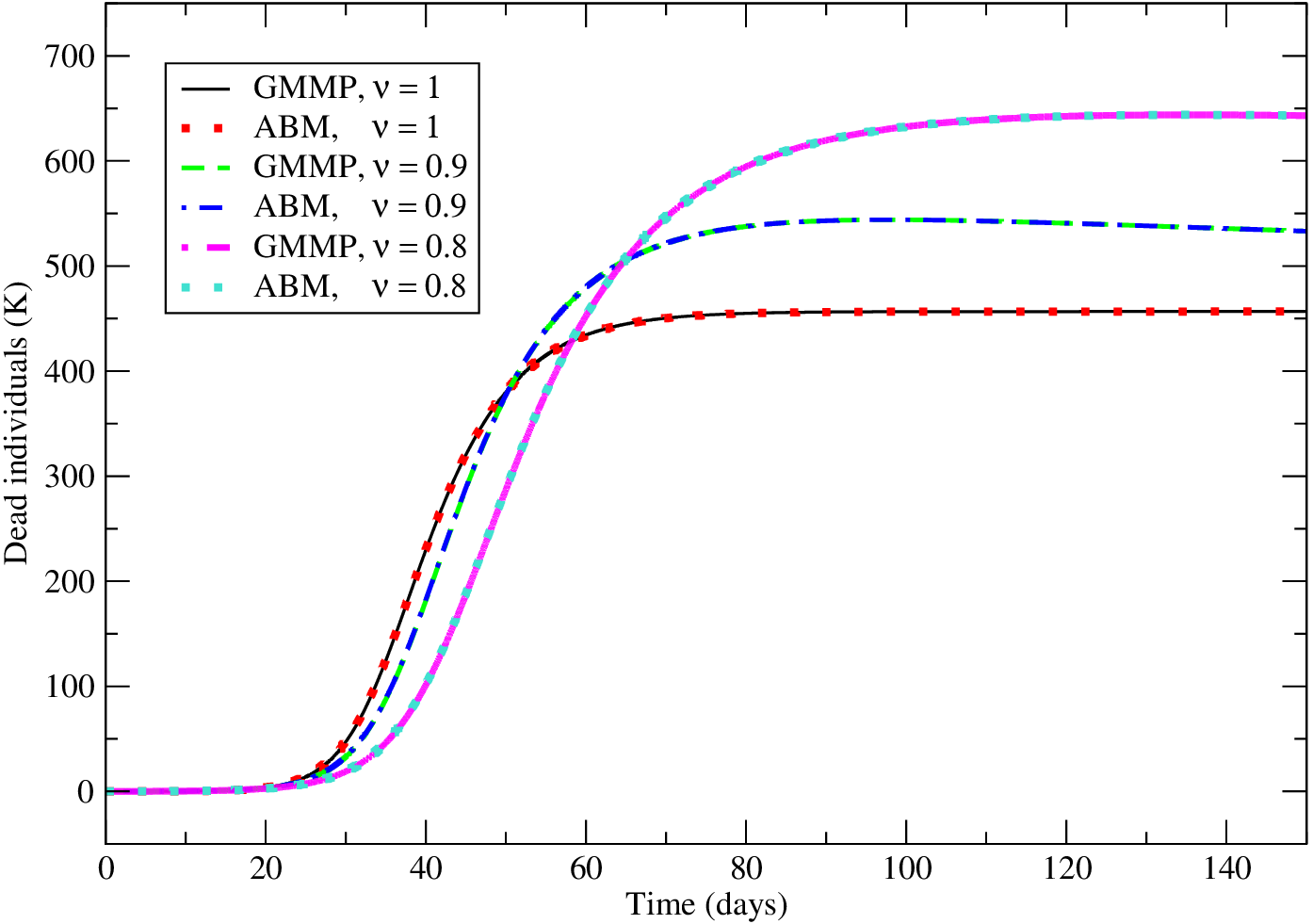}
\vskip0.3cm
\caption{Dead individuals for the classical SEIR model ($\nu = 1$) and fractional-order
derivatives  $\nu = 0.8$ and  $0.9$}
\label{fig5}       
\end{figure}

\begin{figure}
\vskip1cm
\includegraphics[scale=0.35]{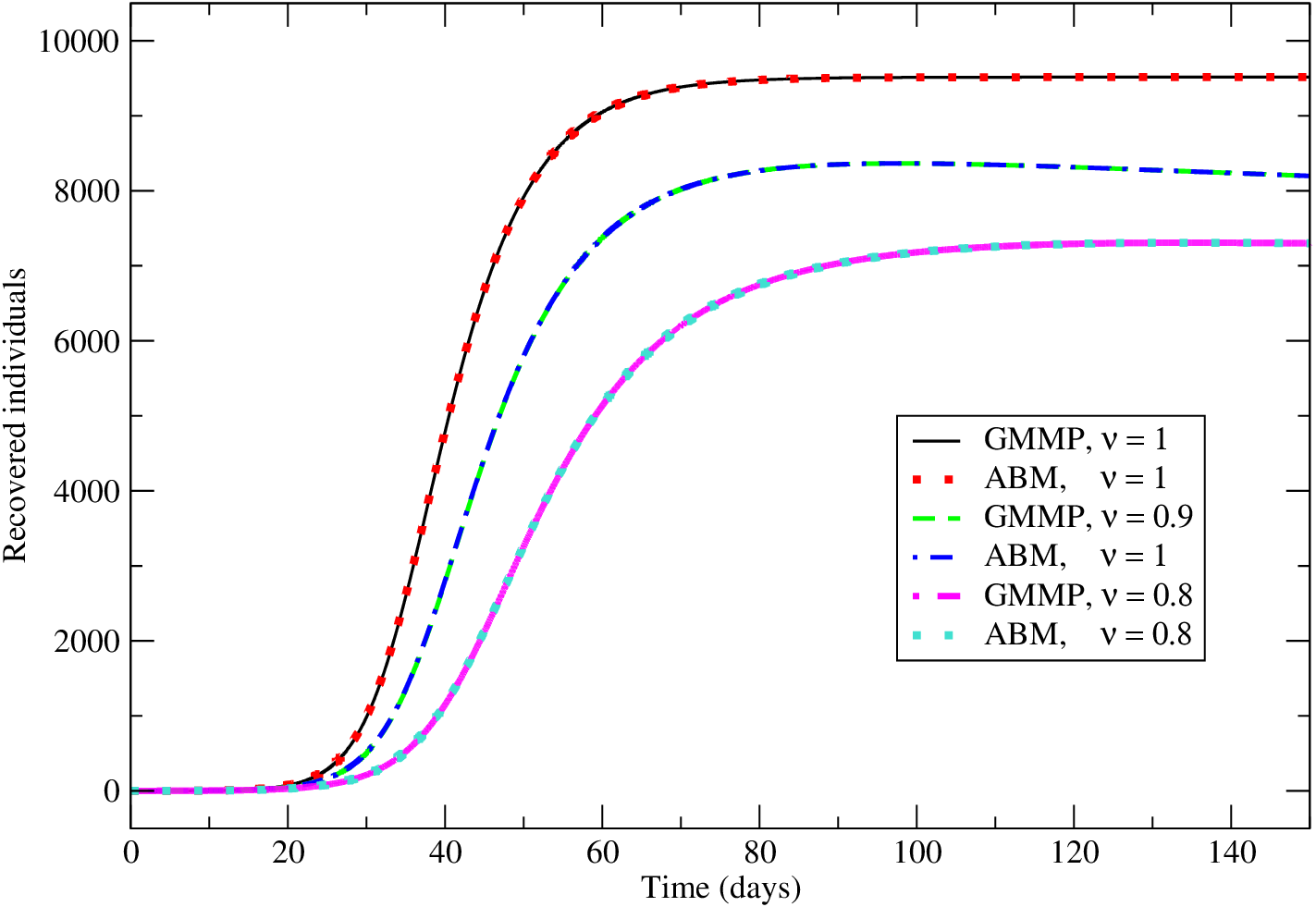}
\vskip0.3cm
\caption{Recovered individuals for the classical SEIR model ($\nu = 1$) and fractional-order
derivatives  $\nu = 0.8$ and  $0.9$}
\label{fig4}       
\end{figure}

\begin{figure}
\vskip1cm
\includegraphics[scale=0.35]{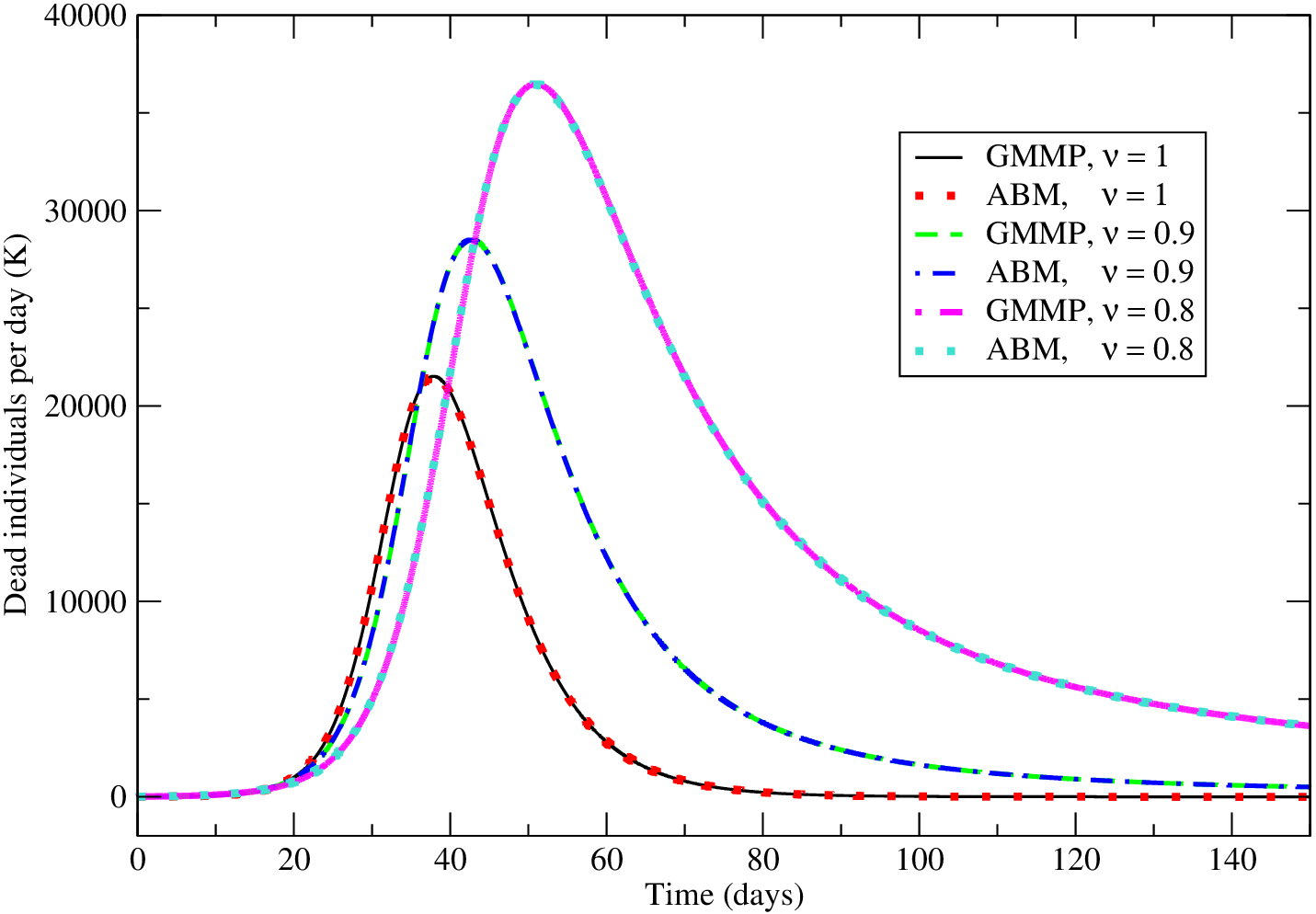}
\vskip0.3cm
\caption{Dead individuals per day for the classical SEIR model ($\nu = 1$) and fractional-order
derivatives  $\nu = 0.8$ and  $0.9$}
\label{fig6}       
\end{figure}

 Figures \ref{fig1}--\ref{fig6} show the results of the  four classes,  S,E,I,R,  and the dead  and dead per day individuals 
 computed by using the GMMP and ABM algorithms. 
 First, an excellent agreement between the results of the two algorithms is observed for all 
 values of the fractional order derivative $\nu$. {\fo To quantify this agreement, we compute  a mean squared relative error 
 between the estimations of  both methods. For example,  in the computation of infected individuals, 
 the following errors are obtained: 1.512 $\times$ 10$^{-5}$ for $\nu = 1$, 9.880 $\times$ 10$^{-6}$ for $\nu = 0.9$ 
 and 1.053 $\times$ 10$^{-5}$ for $\nu = 0.8$.}
 In particular, the  results for $\nu = 1$ agree with those of Figures 1 and 2 in \cite{carcione2020}. 
 Figure \ref{fig1} shows that decreasing the order of the  fractional derivative causes  a delay and an increase in the number of susceptible individuals. While for the classical model the number of infectious individuals vanish  at long times,  this is not the case for the orders  $\nu = 0.8$  and $\nu = 0.9$ (Figure \ref{fig3}). 
We run the simulator up to a very long time but the individuals do not vanish, so that 
 the epidemic never ends (in theory).  This happens because $R_0 \ge 1$. We run other examples with different parameters  such that   $R_0 < 1$ and as expected  the number of infectious individuals vanish and the epidemic dies out.  For brevity these plots are not shown. {\fo The  case  $R_0 < 1$ is analyzed in Subsection 4.2,  when simulating the evolution of the epidemic in the RMBA using fractional derivatives. This value of $R_0$  is associated with the strict lockdown imposed 
by the government, with a corresponding decrease in the number of infected individuals.} 
 
Regarding the exposed infected classes (Figures \ref{fig2}-\ref{fig3}),  a decrease in $\nu$ causes  
delays and  reduces the amplitude  of the peaks  of these classes. 
Furthermore, as $\nu$  decreases  the number of casualties increase as seen in Figure \ref{fig5} while   
Figure \ref{fig6} shows  a delay and increase of the peak in the 
number of dead individuals per day.  {\fo Also, note that  Figure \ref{fig4}  shows a delay and decrease in  the number of recovered individuals as 
the order of the fractional derivative decreases.}

These simulations consider a single value of $\beta$, the lockdown parameter. In a realistic case, $\beta$ is a function of time and the procedure is that every time $\beta$ changes, the algorithm has to be fully initialized from the beginning. Changing $\beta$ in the same time loop yields wrong results. This fact has been verified by cross-checking different algorithms and several fractional orders. 
 
 \vskip1cm
 
 \vskip1cm
 
 \vskip1cm
 
 \vskip1cm
 
 \vskip1cm
 
 \vskip1cm
 
\subsection{Analysis of the COVID-19 epidemic in the RMBA} \label{sec:rmba}

We  model the COVID-19 epidemic in the RMBA, with a population $N_0$ = 14839026 individuals 
according to the 2010 Census (\url {https://www.indec.gob.ar/indec/web/Nivel4-Tema-2-41-135}).
The prediction of the time evolution of the epidemic is very difficult due to the uncertainty 
of the parameters defining the SEIR model. Virus properties such as the infectious  
and incubation periods ($\gamma^{-1}$ and $\epsilon^{-1}$) and life expectancy 
of an infected individual ($\alpha^{-1}$)  
lie in certain  bounded intervals.  Instead,  the parameter $\beta$ is time dependent, due to changes according to  
the lockdown    
and social-distance measures imposed by the government.   
Most authors use the infectious individuals to calibrate the model, e.g., Gonz\'alez-Parra et al. \cite{gonzalez2014}, who model the AH1N1/09 influenza epidemic in Bogot\'a, Colombia 
and  in the Nueva Esparta state in Venezuela. 

Since the number of asymptomatic, undiagnosed infectious individuals in RMBA is unknown,  we 
choose to calibrate the model with the number 
of officially reported casualties as the  most reliable data, from  day 1 (March 9, 2020) 
to day {\fo 198 (September 22th, 2020)} (\url{https://www.argentina.gob.ar/coronavirus/informe-diario}).
Concerning the parameters, fractional order 
and initial  conditions of the model, we assume $\mu$ = 3.6 $\times$ 10 $^{-5}$/ day, corresponding  to 
a life expectancy of  76 years. Changes in the $\beta$ parameter   are associated with different measures of 
lockdown and social distance imposed by the goverment. Thus, we assume that $\beta$ is a piecewise constant 
function, where its variations are related to the inflection points observed in the curve of casualties. 
After  the initial time $t_0 = 1$  day, this curve  shows two  inflection points at times $t_1$ = 31 day  
and  $t_3$ = 50 day. The fractional-order derivative $\nu$,  the  values of $\alpha$, $\beta$, $\epsilon$, 
$\gamma$  and the initial exposed individuals $E(0)$ are estimated by minimizing the $L^2$-norm between the simulated and actual casualties, which is an inverse problem with an infinite number of solutions due  to  the existence of local minima. The estimation is also performed for the  classical case $\nu = 1$.  
This inverse problem is solved by using  a quasi Newton approximation technique for nonlinear least-squares problems, based on the formula 
of Broyden-Fletcher-Goldfarb-Shanno   \cite{Gill81}. Application of this  technique to solve inverse problems in reservoir engineering can be found in \cite{Savioli94}.  Table \ref{table1} shows ranges of 
the fractional derivative $\nu$, of the parameters $\alpha$, $\beta$, $\epsilon$, 
$\gamma$  and the initial  exposed individuals $E(0)$  used in the inversion procedure.
Table \ref{table2} displays the initial values and results of {\fo four}  outputs (Cases) of the  fitting procedure.   

\begin{table}
\caption{Constraints and  ranges of the estimation procedure}
\label{table1}       
\begin{tabular}{lllllll}
\hline\noalign{\smallskip}
Variable $\rightarrow$ &$\nu$ & $\alpha$       & $\beta$      & $\epsilon^{-1}$ & $\gamma^{-1}$ & $E(0)$  \\
                       &      & day$^{-1}$ & day$^{-1}$ &  day        &  day\\
\noalign{\smallskip}\hline\noalign{\smallskip}\\
 Lower bound           & 0.8  & 10$^{-5}$    & 0.1    & 3              &   3     &      10$^2$\\
Upper bound            & 1.0  & 10$^{-1}$    & 0.9       & 9             &   9     &       10$^4$  \\
 \noalign{\smallskip}\hline\noalign{\smallskip}\\
\end{tabular}
\end{table}

\begin{table}
\caption{Initial values and results of the estimation procedure.}
\label{table2}       
\begin{tabular}{lllllllll}
\hline\noalign{\smallskip}
Variable $\rightarrow$ &$\nu$ & $\alpha$             & $\beta_1$       & $\beta_2$         & $\beta_3$     & $\epsilon^{-1}$ & $\gamma^{-1}$& $E(0)$ \\
                       &      & day$^{-1}$       & day$^{-1}$  & day$^{-1}$    &day$^{-1}$ &  day          &  day  &           \\                     
\noalign{\smallskip}\hline\noalign{\smallskip}\\
Case 1   &  &  &  &  &  &  &  &\\
Initial  & 0.9  &6.00$\times$10$^{-3}$   & 0.5              & 0.2                & 0.3    & 5.0     & 4.0  & 500\\
Optimum               & 0.919 &2.130761$\times$10$^{-4}$   & 0.66090              & 0.12507               &0.34002             & 8.976007               & 5.335143         &  1623      \\
$R_0$                  &      &                             & 3.178                  & 0.688                   & 1.725              &                         &                  &           \\
IFR =0.197            &      &                             &                        &                         &                     &                         &                  &           \\
\noalign{\smallskip}\hline\noalign{\smallskip}\\
Case 2   &  &  &  &  &  &  &  &\\
Initial  & 0.85  &6.00$\times$10$^{-3}$   & 0.4              & 0.2                & 0.3    & 5.0     & 4.0  & 1000\\
Optimum                &0.812 &4.179268$\times$ 10$^{-4}$&0.77273                &0.47231                 & 0.56801              &8.121503                &3.022527          & 1138   \\
$R_0$                  &      &                              & 1.982                  & 1.329                 &  1.539     &                  &            &       \\
IFR =  0.444            &      &                              &              &            &            &                  &            &       \\
\noalign{\smallskip}\hline\noalign{\smallskip}\\
Case 3   &  &  &  &  &  &  &  &\\
Initial  & 1  &6.00$\times$10$^{-3}$   & 0.5              & 0.2                & 0.3    & 5.0     & 4.0  & 500\\
Optimum                & 1    &2.822018 $\times$10$^{-4}$    & 0.49040     & 0.10396   & 0.27568   & 8.975264         &6.212071    &  2821\\
$R_0$                  &      &                              & 3.041         & 0.645      & 1.710      &                  &            &       \\
IFR = 0.175            &      &                              &              &            &            &                  &            &       \\
\noalign{\smallskip}\hline\noalign{\smallskip}\\
Case 4   &  &  &  &  &  &  &  &\\
Initial  & 0.9  &6.00$\times$10$^{-3}$   & 0.4              & 0.2                & 0.3    & 5.0     & 4.0  & 1000\\
Optimum                & 0.929   &2.787611 $\times$10$^{-4}$    & 0.47289     & 0.10168   & 0.31122   & 8.244641         &5.751017    &  4110\\
$R_0$                  &      &                              & 2.526         & 0.606      & 1.713      &                  &            &       \\
IFR = 0.254            &      &                              &              &            &            &                  &            &       \\
\noalign{\smallskip}\hline\noalign{\smallskip}\\
\end{tabular}
\end{table}




Let us analyze {\fo four} cases, resulting from the 
minimization algorithm. We obtained the SEIR parameters, the fractional order and 
the initial exposed humans values fitting the data. 
In all the cases, the initial number of infected individuals is  
assumed to be $I(0)$ = 100. 

\begin{figure}
  \vskip1cm
\includegraphics[scale=0.35]{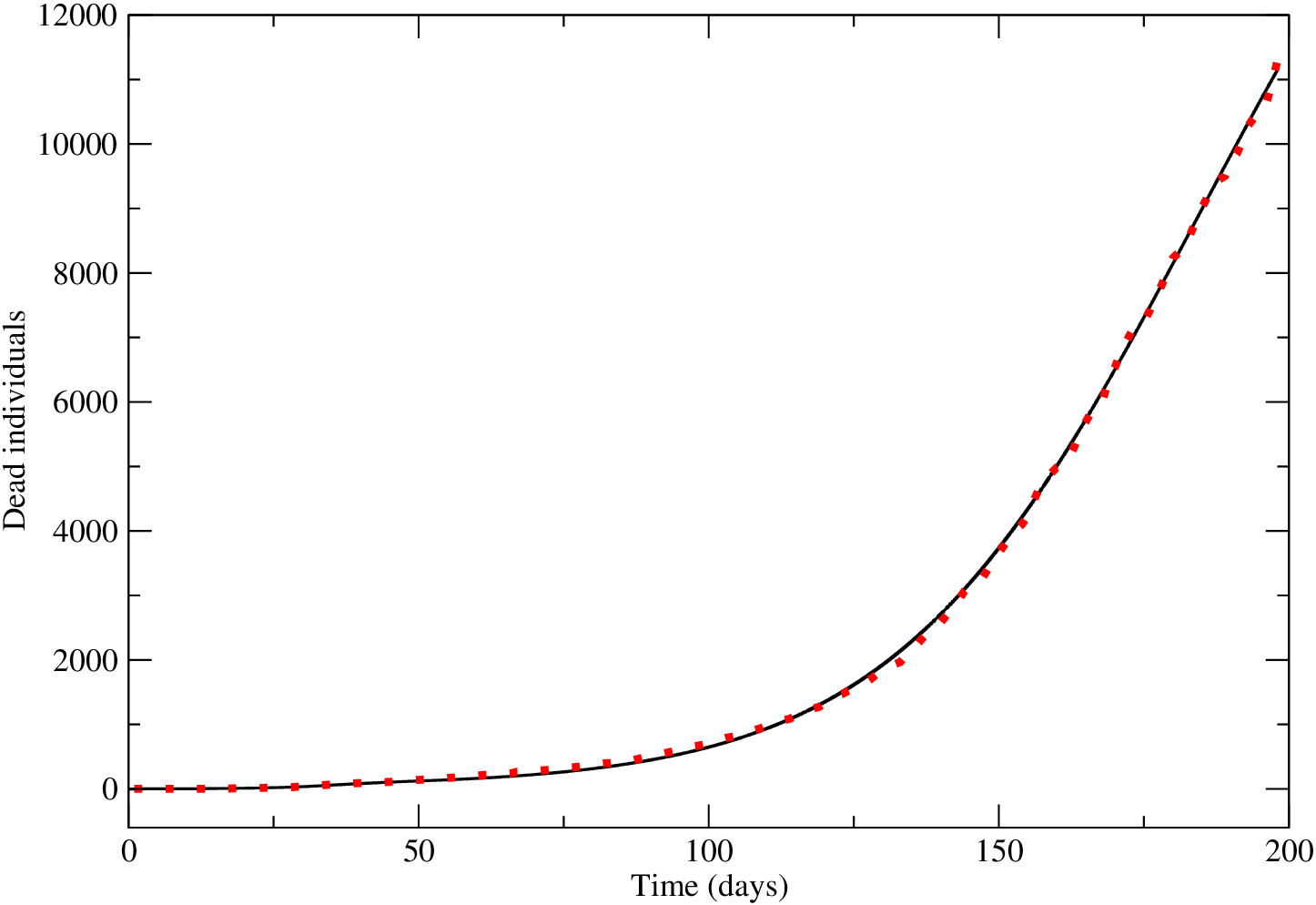}
\vskip0.3cm
\caption{Dead  individuals. The red dots represent the data and the solid line the fit using 
 the  SEIR model of fractional order with $\nu = 0.919$}
\label{fig7}       
\end{figure}

\begin{figure}
\vskip1cm
\includegraphics[scale=0.35]{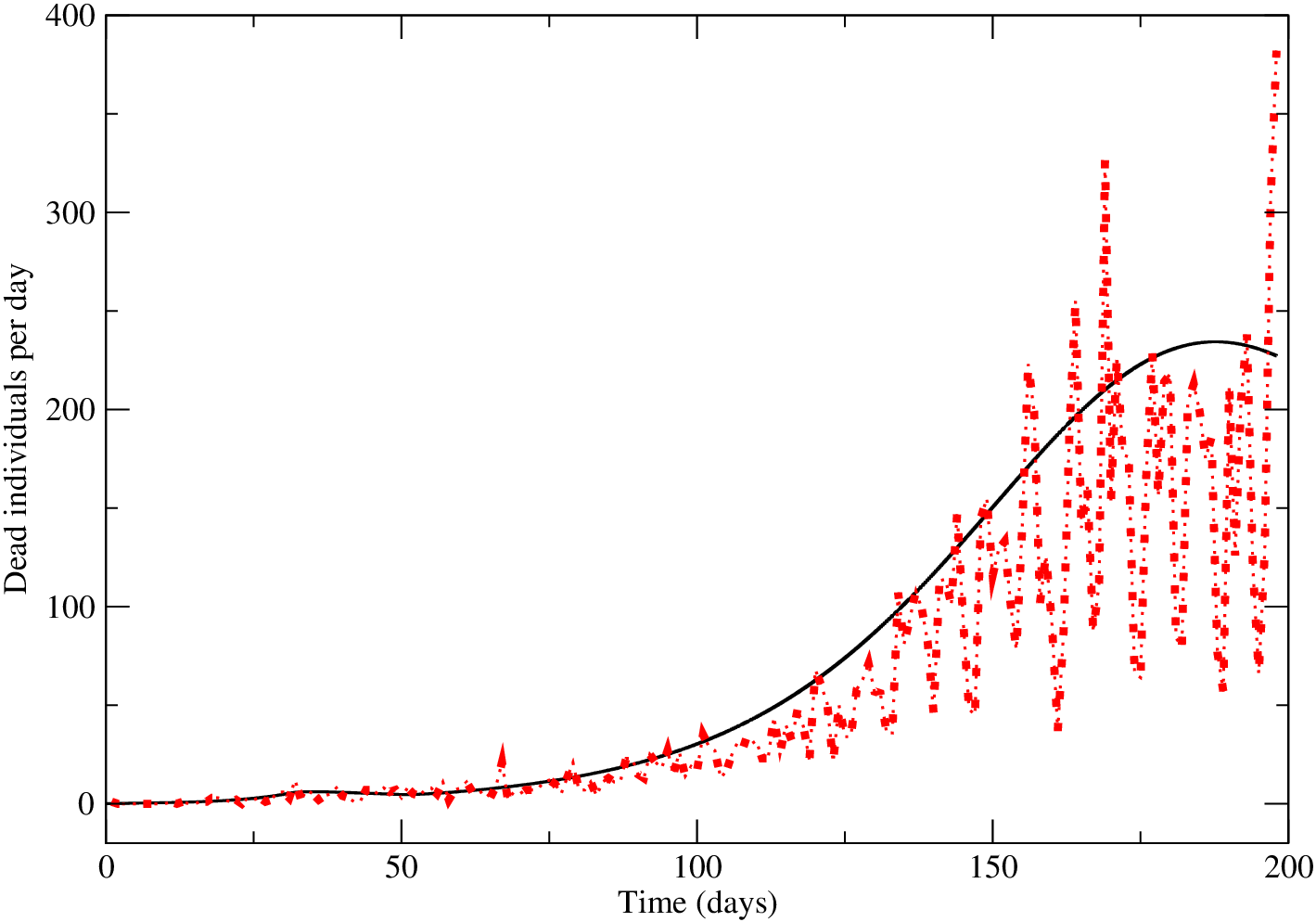}
\vskip0.3cm
\caption{Dead  individuals per day. The red dots represent the data and the solid line the fit using 
 the  SEIR model of fractional order with $\nu = 0.919$}
\label{fig8}       
\end{figure}

\begin{figure}
\vskip1cm
\includegraphics[scale=0.35,angle=0]{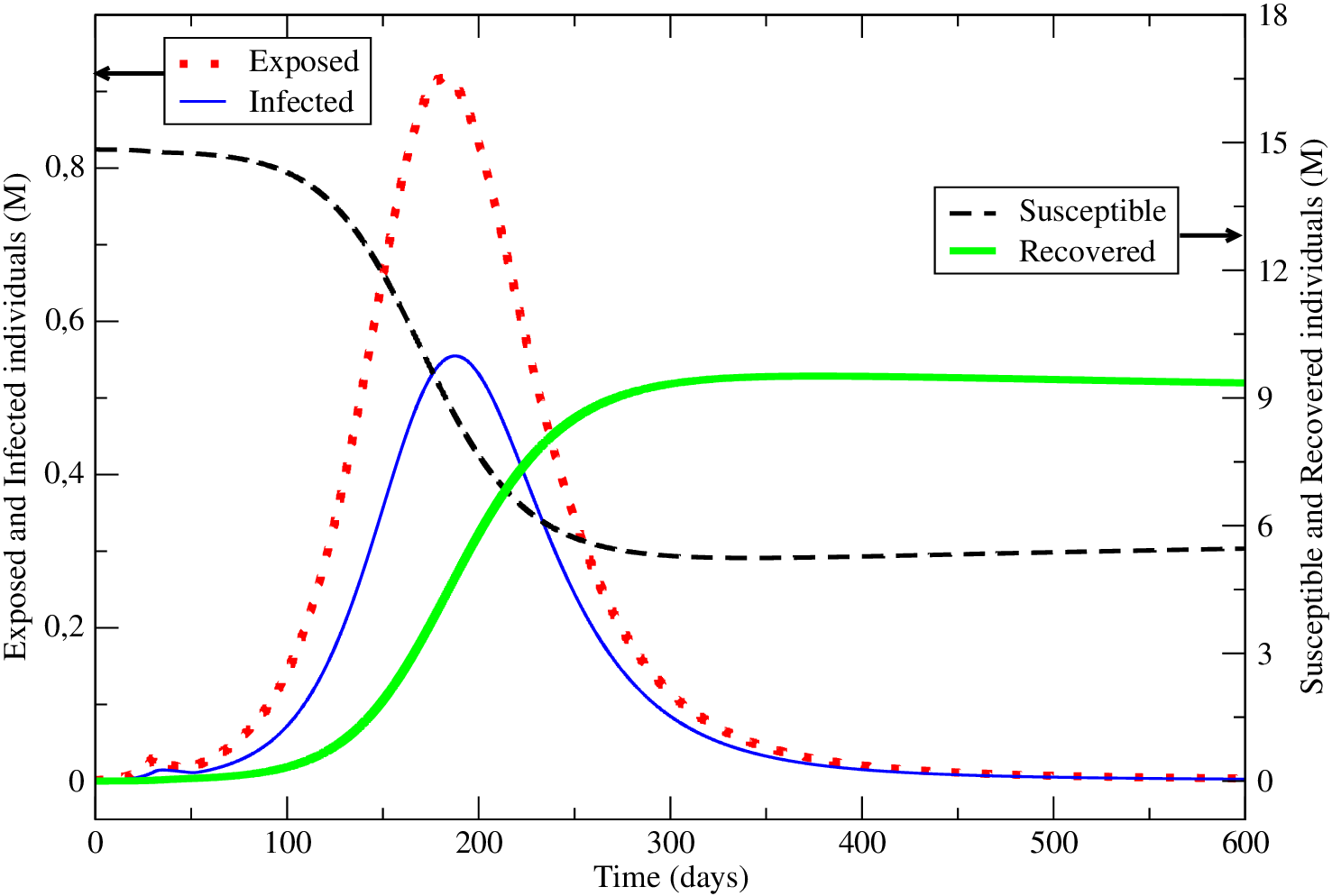}
\vskip0.3cm
\caption{Number of individuals in all classes (millions) for the  SEIR model of fractional order with $\nu = 0.919$}
\label{fig9}       
\end{figure}

\begin{figure}
\vskip1cm
\includegraphics[scale=0.35,angle=0]{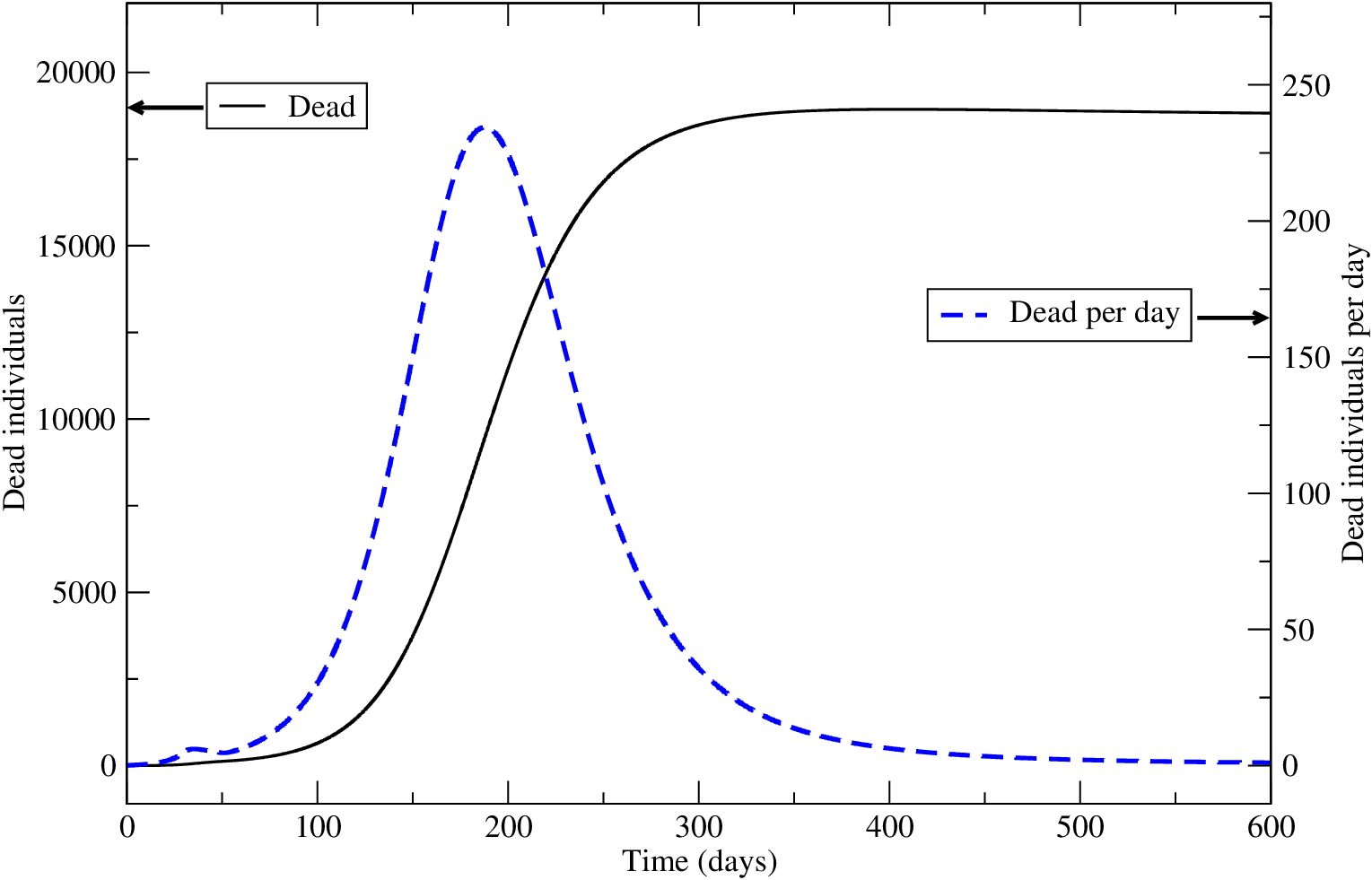}
\vskip0.3cm
\caption{Total number of deaths and deaths per day for the  SEIR model of fractional order with $\nu = 0.919$}
\label{fig10}       
\end{figure}

{\fo Figures \ref{fig7} and \ref{fig8}  show the dead individuals and dead individuals per day for  Case 1. The inflection point at $t_1 = 30$ day, 
related to a change of $R_0$ from  3.178 
to  0.688, shows  
  a decay in the  simulated curves, because of the effect of the lockdown. 
  After $t_1 = 50$ day, the curves  exhibits a continuous
  increase in casualties due to the relaxation of the lockdown measures with  $R_0 = 1.725$.
  Figure  \ref{fig9}  shows the behavior of all classes, with a  
  a peak of  555 thousand  infected  individuals at day 188 (September 12th, 2020) while Figure \ref{fig10} 
exhibits  a  death toll of 19000 people after 800 days (May 17th, 2022) 
and a peak of 234 casualties at day 188.
  
The parameters of Cases 2 and 3 in Table \ref{table2} also fit the data, with  graphs similar to those in 
Figures \ref{fig7} and \ref{fig8}. Case 2 
estimates peaks of 309 deaths  and  285 thousand  infected individuals
at day 222 (October 16th, 2020). At day 800 (May 17, 2022), 
there are 34 thousand deaths and 7457 thousand  recovered humans.   
 This increase in the 
number of casualties is due to the higher infection fatality
rate IFR and higher reproduction ratios $R_0$ as compared with those of Case 1 (see
Table 2).
 
Case 3, which corresponds to the classical SEIR model ($\nu$ = 1), exhibits a peak of 171 casualties at day 184 (September 8th, 2020) and 
607 thousand  people infected.
The end of the epidemic is consider 
the day  at which  the number of infected individuals  is smaller than 1,  which is day 594 (October 24th, 2021) 
for this case. 
At this day,  the total number of  recovered  and dead individuals 
are 10157 thousand  and 
18 thousand, respectively, so that the total number of infected people at the end of the epidemic is 
10175  thousand individuals. This is the case 
predicting the smallest number of casualties.}

{\fo Finally, since the reported number of deceased people could possibly be underestimated due to undeclared cases
and delays in the upload of official  data, we also consider a case with
30 \% more casualties to date (Case 4 in Table \ref{table2}), giving IFR = 0.254 \% and values
of the parameters similar to those of Case 1. Besides, the peak occurs almost at the same day of Case 1 
(day 187: September 11th, 2020) with 592 thousand  infected  individuals and 296 casualties. This peak of casualties 
and the death toll of 24400 individuals are
approximately 30 \% higher than those of Case 1.}

\vskip1cm

\vskip1cm

\vskip1cm

\begin{figure}
\vskip1cm
\includegraphics[scale=0.35,angle=0]{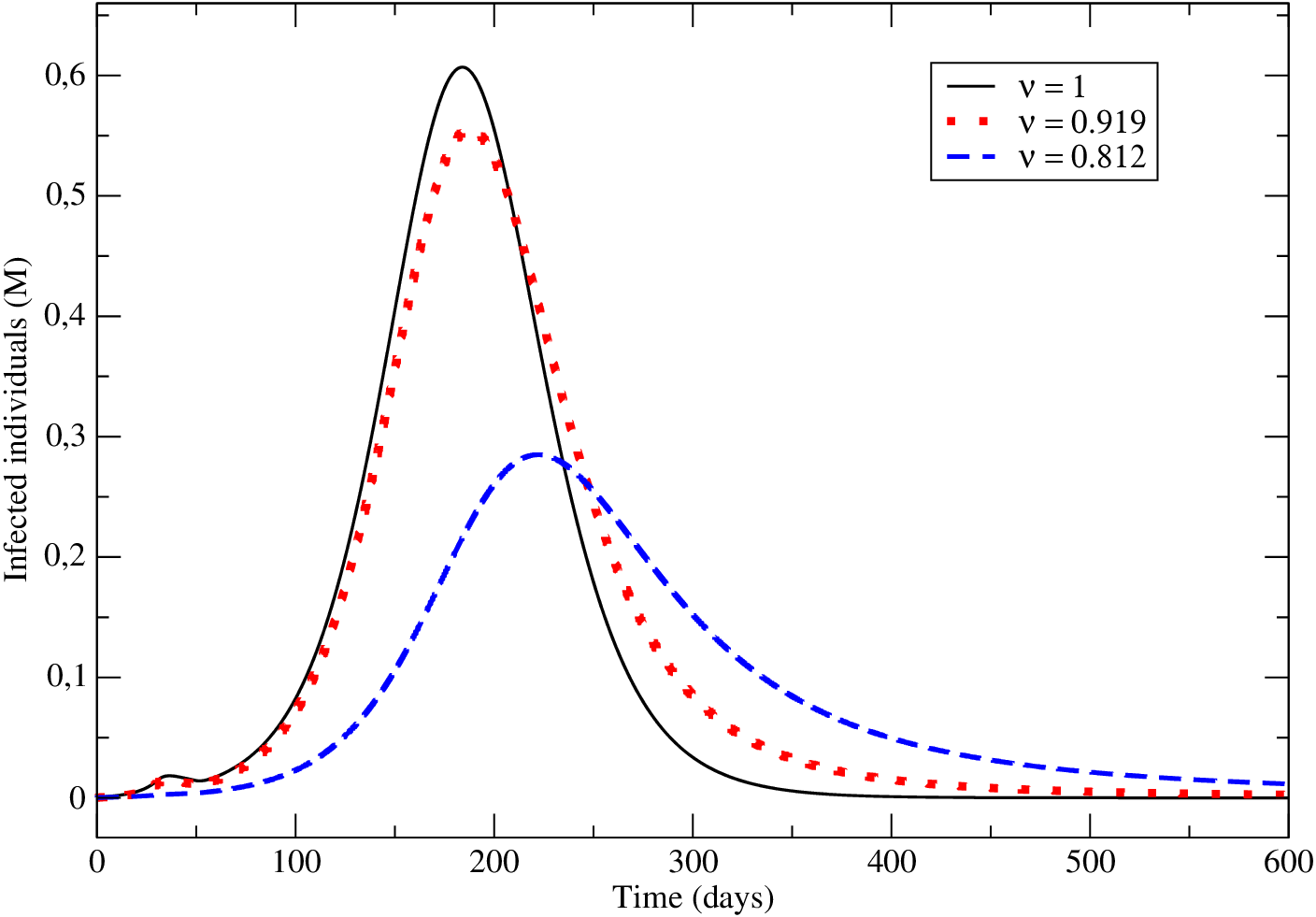}
\vskip0.3cm
\caption{Infected individuals for the   SEIR model of fractional orders $\nu =1, 0.919$ and  $0.812$}
\label{fig11}       
\end{figure}

\begin{figure}
\vskip1cm
\includegraphics[scale=0.35,angle=0]{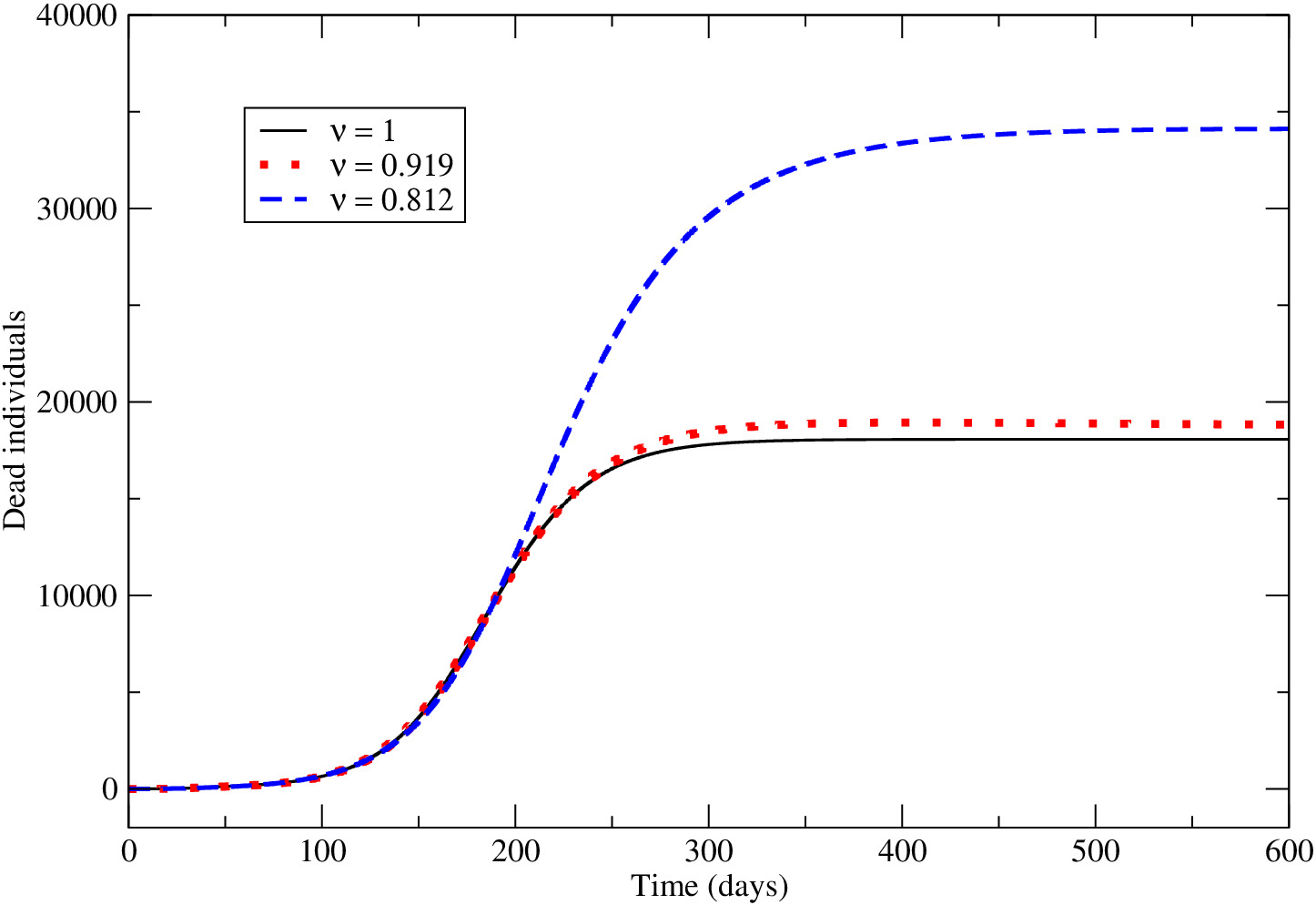}
\vskip0.3cm
\caption{Dead individuals for the   SEIR model of fractional orders $\nu =1, 0.919$ and  $0.812$}
\label{fig12}       
\end{figure}

\begin{figure}
\vskip1cm
\includegraphics[scale=0.35,angle=0]{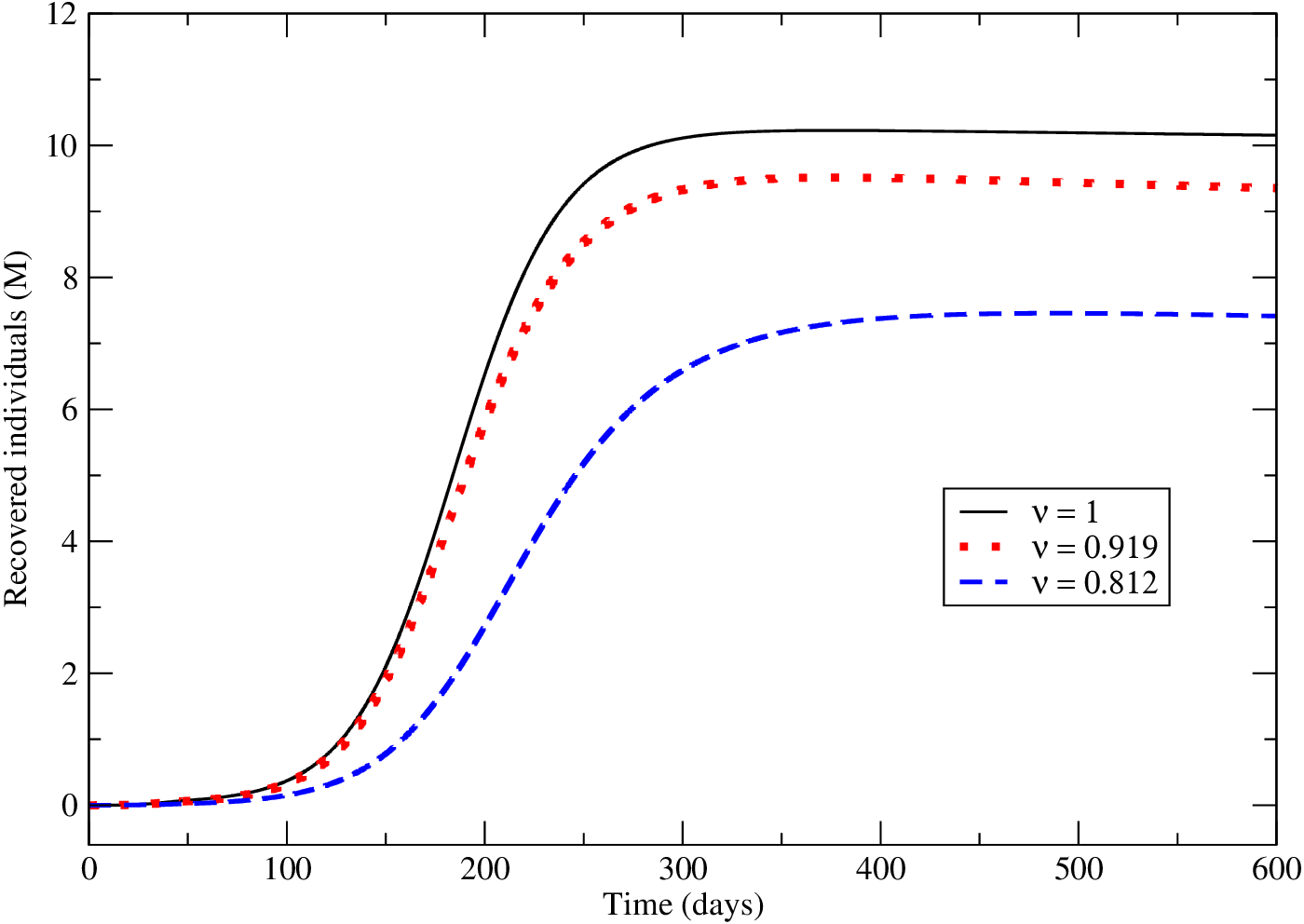}
\vskip0.3cm
\caption{Recovered  individuals for the   SEIR model of fractional orders $\nu =1, 0.919$ and  $0.812$}
\label{fig13}       
\end{figure}

\begin{figure}
\vskip1cm
\includegraphics[scale=0.35,angle=0]{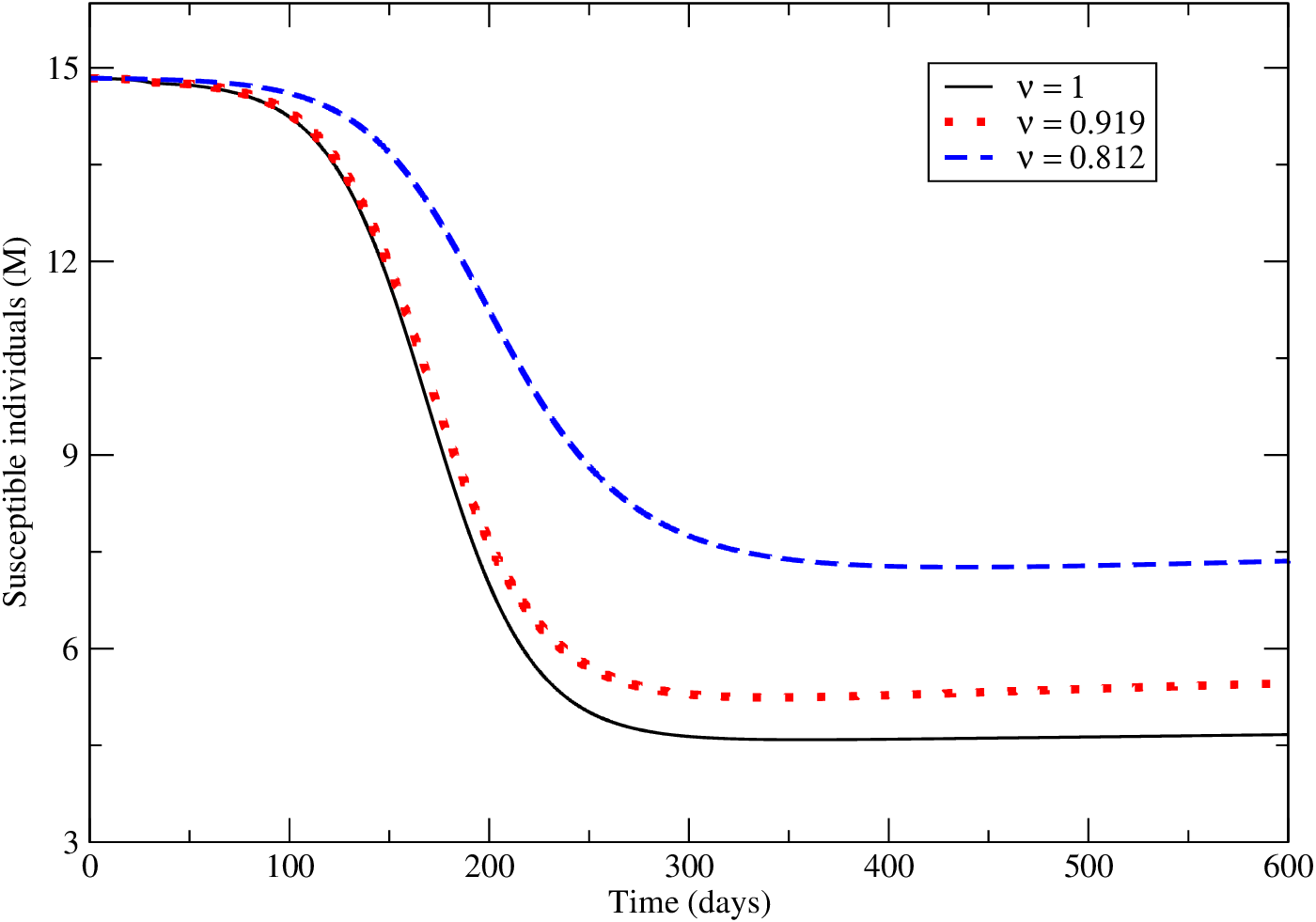}
\vskip0.3cm
\caption{Susceptible individuals for the   SEIR model of fractional orders $\nu =1, 0.919$ and  $0.812$}
\label{fig14}       
\end{figure}

{\fo In the following, we compare the behavior of all classes for the different orders of the fractional derivative
used in this analysis, i.e., $\nu = 1, 0.919$ and $0.812$.
Figure \ref{fig11} displays the number of infected individuals, where there is a delay and decrease 
of the peak values as the order of the fractional derivative decreases. This behavior  
is consistent with that observed in  Figure \ref{fig3}. 
Figure \ref{fig12}  shows an increase in the number of  casualties by decreasing the order 
of the fractional derivative, with a 47 \% increase  between  $\nu = 1$  and $\nu = 0.812$. Moreover, it
can be seen  that the curves stabilize at later times as the fractional order decreases.
Finally, Figures \ref{fig13} and \ref{fig14} exhibit the estimated recovered and susceptible individuals for the three values of $\nu$. 
Recovered individuals increase and, consequently, susceptible individuals decrease  as the order of fractional derivative increases.
The curves exhibit asymptotic values at later times as $\nu$ decreases, and the lower the value of $\nu$ 
the later individuals recover from the virus infection. 
Note that the general trends of Figures \ref{fig11}--\ref{fig14} are similar to those of the figures in Subsection  \ref{vali}, 
in spite of the fact that parameters obtained from the adjustment are different for the three cases.}

{\fo In the four cases described above, we consider that  the initial number of infected individuals is $I(0)= 100$. Nevertheless, we tested other values:
if $I(0)$ belongs to the interval $[10, 150]$ a reasonable adjustment is obtained, with similar values to those shown in Table \ref{table2} and a slight delay 
on the infected individuals peak as  
$I(0)$ decreases. Outside this interval, the fit is poor and the results have no physical meaning.}

\vskip1cm

\vskip1cm

\section{Conclusions}

{\fo We use a fractional SEIR  (Susceptible, Exposed, Infected, Recovered)  diffusion model to analyze the evolution of the COVID-19 epidemic in Argentina, particularly in the Region Metropolitana de Buenos Aires (RMBA), where a significant number of the population is concentrated.

We solve the SEIR  system of fractional order $\nu, 0< \nu< 1$ and   the classical ($\nu = 1$) SEIR model by using a time-explicit  
Gorenflo-Mainardi-Moretti-Paradisi (GMMP) method.  To validate this method, the results were  cross-checked  with those of the time-explicit  fractional Adams-Bashford-Moulton (ABM) method, obtaining an excellent agreement between the two schemes.

Assuming that the birth and death rates are balanced, the parameters that characterize the model are the infection and incubation periods, the probability of disease transmission per contact, the fatality rate and the initial number of exposed individuals. 
These parameters and the order $\nu$ of the fractional derivative are estimated by fitting the number of casualties officially reported. This inverse problem is solved by using a quasi-Newton technique for non-linear least-squares problem with the Broyden-Fletcher-Goldfarb-Shanno formula.

In all the simulations  we used  three lockdown parameters (denoted by $\beta$),  
associated with the different  
measures taken by the government during the evolution of the epidemic. 
One important conclusion  related with this time-dependent parameter is that both the fractional GMMP and  ABM algorithms need to be fully initialized from  the beginning in order to obtain correct results.

Different cases have been analyzed since the inverse problem has an infinite number of solutions. We observe a similar behavior in all the cases, with a fatality rate IFR varying in the range, $[0.175 , 0.444]$. After the 50th day of lockdown, it is observed a continuous increase in casualties 
due to the relaxation of the preventive social isolation and community circulation of the virus.

The numerical simulations in RMBA show that  when  the order of the fractional derivative  decreases, 
i.e., higher subdiffusion  of the virus, 
 the duration of the epidemic is extended, and the peak of infected individuals and number of casualties increase. Furthermore, the classical SEIR model  yield a smaller number of casualties and infected 
 individuals with associated peaks located at earliest  times as compared with those of the fractional-order cases.
 }
 
\section{Appendix}

The Adams-Bashford-Moulton explicit scheme for the fractional order SEIR equations is formulated as follows \cite{lizeng2015}
\vskip0.3cm
\noindent
{\bf Predictor}
\begin{eqnarray}
&&S^p_{n+1} = ((n+1) \Delta t)  S_0 + \sum_{j=0}^n b_{j,n+1)} f_1^{\nu}(S_j, E_j, I_j, R_j)\label{abm1}\\
&&E^p_{n+1} = ((n+1) \Delta t)  E_0 + \sum_{j=0}^n b_{j,n+1)} f_2^{\nu}(S_j, E_j, I_j, R_j)\nonumber\\
&&I^p_{n+1} = ((n+1) \Delta t)  I_0 + \sum_{j=0}^n b_{j,n+1)} f_3^{\nu}(S_j, E_j, I_j, R_j)\nonumber\\
&&E^p_{n+1} = ((n+1) \Delta t)  R_0 + \sum_{j=0}^n b_{j,n+1)} f_4^{\nu}(S_j, E_j, I_j, R_j)\nonumber\\
&&N^p_{n+1} = S^p_{n+1} +  E^p_{n+1} + R^p_{n+1} + I^p_{n+1}.\nonumber 
\end{eqnarray}
{\bf Corrector}
\begin{eqnarray}
&&S_{n+1} = ((n+1) \Delta t)  S_0 
+ \sum_{j=0}^n a_{j,n+1}  f_1^{\nu}(S^p_{n+1}, E^p_{n+1}, I^p_{n+1},  R^p_{n+1}) \label{abm2}\\
&&E_{n+1} = ((n+1) \Delta t)  E_0 
+ \sum_{j=0}^n a_{j,n+1}  f_2^{\nu}(S^p_{n+1}, E^p_{n+1}, I^p_{n+1},  R^p_{n+1}) \nonumber\\
&&I_{n+1} = ((n+1) \Delta t)  I_0 
+ \sum_{j=0}^n a_{j,n+1}  f_3^{\nu}(S^p_{n+1}, E^p_{n+1}, I^p_{n+1},  R^p_{n+1}) \nonumber\\
&&R_{n+1} = ((n+1) \Delta t)  R_0 
+ \sum_{j=0}^n a_{j,n+1}  f_4^{\nu}(S^p_{n+1}, E^p_{n+1}, I^p_{n+1},  R^p_{n+1}) \nonumber\\
&&N_{n+1} = S_{n+1} +  E_{n+1} + R_{n+1} + I_{n+1}.\nonumber
\end{eqnarray}
In \eqref{abm1}-\eqref{abm2} the coefficients $b_{j,n+1},  a_{j,n+1}$ are 
\begin{eqnarray*}
&&b_{j,n+1} = \dfrac{1}{\Gamma(1 + \nu} \left[(n -j +1)^\nu -  (n -j)^\nu \right]\\
&& a_{j,n+1} = \dfrac{1}{\Gamma(2 + \nu)}
=\begin{cases} (n)^{\nu +1} - (n-\nu) (n+1)^\nu, \quad j=0,\\
 (n - j +2)^{\nu+1} + (n - j)^{\nu+1}  - 2 (n-j+1)^{\nu+1},\quad 1\le j \le n-1\\
1 , \quad j=n+1.\end{cases} 
\end{eqnarray*}

{\fo Concerning the error of the numerical scheme ABM, Abdullah et al. \cite{abdullah2017}
give a bound in terms of the time  step size $\Delta t$. 
On the other hand, Li and Zeng \cite{lizeng2015}  and  Li et al. \cite{li2011}  show  that the fractional  forward Euler and ABM methods  are stable and  convergent of order one in  $\Delta t$. }


\end{document}